\documentclass{elsart}
\usepackage[dvips]{graphics}
\begin{document}
\bibliographystyle{unsrt}

\journal{Astroparticle Physics}

\newcommand{\mocca} {{\it MOCCA}}
\newcommand{\corsika} {{\it CORSIKA}}
\newcommand{\sibyll} {{\it SIBYLL}}
\newcommand{\qgsjet} {{\it QGSjet}}
\newcommand{\mocspl} {{\it MOCCA-Internal}}
\newcommand{\mocsib} {{\it MOCCA-SIBYLL}}
\newcommand{\corsib} {{\it CORSIKA-SIBYLL}}
\newcommand{\corqgs} {{\it CORSIKA-QGSjet}}
\newcommand{\geisha} {{\it GEISHA}}
\newcommand{\egs} {{\it EGS4}}
\newcommand{\nm} {\ensuremath{N_{max}}}
\newcommand{\xm} {\ensuremath{X_{max}}}
\newcommand{\xr} {\ensuremath{X_{rise}}}

\begin{frontmatter}
\title{A Comparative Study of the Depth of Maximum of Simulated Air
Shower Longitudinal Profiles}
\author{C.L. Pryke}
\address{University of Chicago, The Enrico Fermi Institute,
              5640 S. Ellis Avenue, Chicago, Illinois 60637-1433, USA.
              email: pryke@aupc1.uchicago.edu}

%%%%%%%%%%%%%%%%
\begin{abstract}
A comparative study of simulated air shower longitudinal profiles
is presented.
An appropriate thinning level for the calculations is first
determined empirically.
High statistics results are then provided, over a wide energy range,
($10^{14.0}$ to $10^{20.5}$~eV), for proton \& iron primaries,
using four combinations of the \mocca\ \& \corsika\ program
frameworks, and the \sibyll\ \& \qgsjet\ high energy hadronic
interaction models.
These results are compared to existing experimental data.
The way in which the first interaction controls \xm\ is investigated,
as is the distribution of \xm.
\end{abstract}

%%%%%%%%%%%%%%%%

\end{frontmatter}

%%%%%%%%%%%%%%%%%%%%%%%%%%%%%%%%%%%%%%%%%%%%%%%%%%%%%%%%%%%%%%%%%%%%%%%%%%%%%%%
\section{Introduction}

% E spectrum, direct measure limit & comp, SN cutoff, high E mystery
The energy spectrum of cosmic rays is a power law with the flux
falling by three orders of magnitude for each decade increase in
energy.
At $\sim 10^{14}$~eV the flux becomes so low that
current balloon and satellite experiments lack the exposure required
to detect a significant sample of events.
This is unfortunate as the nature of the primaries remains
of great astrophysical interest.
Where direct measurements are possible the cosmic rays are
known to be mostly protons and atomic nuclei.
The most plausible acceleration site is at the shock fronts
produced by supernova explosions.
However, theoretical considerations predict a maximum energy from
this process of $\sim 10^{15}$~eV, whereas the energy spectrum is
observed to continue with only small deviations up to $>10^{19}$~eV.
The origin of the particles at $>10^{15}$~eV is somewhat
mysterious.

% Old problem, what is EAS
It has long been supposed that insight would result
if the composition of the primaries could be measured.
Due to the extremely low flux the only way to get information
on these particles is to study the extensive atmospheric cascades
which they initiate.
When a cosmic ray enters the atmosphere it collides with the nucleus
of an air atom, producing a number of secondaries.
These go on to make further collisions, and the number of
particles grows.
Eventually the energy of the shower particles is degraded to
the point where ionization losses dominate, and their number
starts to decline.

% Long prof, smooth curve, Xmax
It is a coincidence that at the energy where direct detection
of the cosmic rays becomes impractical, the resulting air showers
become big enough to be easily detectable at ground level.
The number of particles in the cascade also becomes large enough
that the longitudinal profile, or number of particles versus
atmospheric depth, becomes a smooth curve, with a well
defined maximum.
This maximum depth, referred to as \xm, is often regarded as
the most basic parameter of an air shower, and much effort
has been expended to measure and interpret it.

% Xmax inc with E, given E rel to A, superposition
The depth of maximum increases with primary energy as more
cascade generations are required to degrade the secondary particle
energies.
For given total energy \xm\ is related to the energy per nucleon
of the primary.
To first order the interactions occur between individual nucleons
of the primary, and the target air nuclei.
Therefore a shower initiated by a compound nucleus can be
thought of as the superposition of many proton initiated
showers, with correspondingly lower energy.

% Hard problem, fluctuations, mean values
Unfortunately, of course, the detail is not so simple.
For a number of reasons extracting information on the nature of
the cosmic ray primaries from the air showers they produce
has proved to be exceedingly difficult.
The most fundamental problem is that the initial interactions are
subject to large inherent fluctuations.
This limits the event-by-event mass resolution of even an ideal
detector.
However, progress can still be made by looking at mean parameter
values, or better, their distributions.

% Uncertain models required
The second major problem is that sophisticated modeling is
required to predict the absolute value of an observable parameter
which is expected for a primary of given type and energy.
Nucleus-nucleus interactions at the energies of the first few cascade
steps are well beyond the reach of accelerator experiments.
Therefore it is necessary to rely on hadronic interaction models
which attempt to extrapolate from the available data using different
mixtures of theory and phenomenology.
The lower energy part of the cascade can be modeled using
well known physics, although the programs are complex with
corresponding scope for errors.

% Xmax expt, poor res and stat
The depth of shower maximum has been determined by a number of experiments.
In the energy range $10^{14}$ to $10^{16}$~eV it has been measured
with varying degrees of directness using \v{C}erenkov
light~\cite{dice,airobicc,blanca}.
The range $10^{17}$ to $10^{19}$~eV has been observed rather directly
by the Fly's Eye detector through fluorescence light~\cite{Bird93_1}.
Finally the region above $10^{19}$~eV is the focus of the HiRes
Fly's Eye~\cite{HiRes}, Auger Project~\cite{Auger}, and others.
In the past the experimental resolution and statistics have often
been so poor that the mean value of \xm\ has been discussed 
rather than its distribution --- this is changing.

% Need for consistent sims over wide E range
The simulations required to interpret the data from any given
experiment have usually been performed only for the energy
range accessible to it.
This is unfortunate since checking the consistency of experiments
in adjacent energy ranges is critically important, given the
uncertainty of the high energy hadronic interaction models.
Additionally the exact value of \xm\ for a given model can
depend on the way in which the longitudinal profiles are
recorded and fit.
The purpose of this paper is to provide \xm\ values with good
statistical precision, over a wide energy range, and computed
in a consistent way using several hadronic models and two
different cascade ``framework'' programs; for a more detailed
discussion see~\cite{GAP-98-035}.

% Framework can call 3rd party, MOCCA/CORSIKA, SIBYLL/QGSjet.
The process of air shower simulation can be broken up into
several parts.
A framework program is required which handles the mechanics
of the process and calls appropriate subroutines to
model the interaction and propagation of the particles.
Some fraction of the required transport and interaction modeling
may be provided using third party code.
In this paper two air shower simulation packages which have been
heavily used in the literature are considered.
The first is \mocca\ written by Hillas~\cite{MOCCA}.
This originally used a simple, built-in hadronic interaction model,
but has also been linked to \sibyll~\cite{Sibyll};
all other modeling is handled internally.
The second program is \corsika, a well documented and thorough
program prepared originally for the Kascade experiment~\cite{CORSIKA}.
It is linked to a number of high energy hadronic models,
two of which are suitable for use over the very wide
energy range of this study; \sibyll\ and \qgsjet~\cite{QGSjet}.
An attractive feature of this program is the use of the well
established High Energy Physics codes \egs~\cite{EGS4} and
\geisha~\cite{GEISHA}, for the electromagnetic, and lower energy
hadronic modeling respectively.
See Table~\ref{tab:simprogs} for a summary.

\begin{table}
\begin{tabular}{|l|l|l|l|l|}
\hline
Program frame & High E Hadronic & Low E Hadronic & Electromagnetic \\
\hline
\mocca & Internal & Internal & Internal \\
       & \sibyll && \\
%MOCCA & $\left( \begin{array}{ll}Internal\\SIBYLL\end{array}\right)$
\hline
\corsika & \sibyll & \geisha & \egs \\
         & \qgsjet && \\
\hline
\end{tabular}
\label{tab:simprogs}
\caption{Summary of the four program-frame / interaction-model combinations.}
\end{table}

Due to the inherent limitations of air shower fluctuations,
and also because of poor experimental resolution and statistics, \xm\ data
is often compared only to simulated values for proton
and iron nuclei primaries.
These are generally regarded as the extreme ends of the
possible range.
At lower energies the composition of cosmic rays tracks the general
abundances of solar system material, with some modifications
due to propagation spallation effects.
Iron is the heaviest significantly abundant element.

%%%%%%%%%%%%%%%%%%%%%%%%%%%%%%%%%%%%%%%%%%%%%%%%%%%%%%%%%%%%%%%%%%%%%%%%%%%%%%%
\section{Technical Details}
\label{sec:detail}

At the highest cosmic ray energies it is absolutely necessary to
use techniques which accelerate the simulation process.
A popular approach is called thinning: below a threshold energy
only a sub-set of the particles are tracked, with weightings to
compensate for those discarded~\cite{MOCCA}.
The threshold is usually specified as a fraction of
the primary energy, and referred to as the ``thinning level''.

For this study \mocca92 and \corsika\ 5.62 were used. 
This version of \corsika\ includes a very similar thinning algorithm
to \mocca.
In all cases the high energy hadronic interaction model was used
with the set of inelastic cross sections provided by its authors.
Electromagnetic particle energy cutoff was a uniform 0.2~MeV.

\begin{figure}
\begin{center}
\resizebox{\textwidth}{!}{\includegraphics{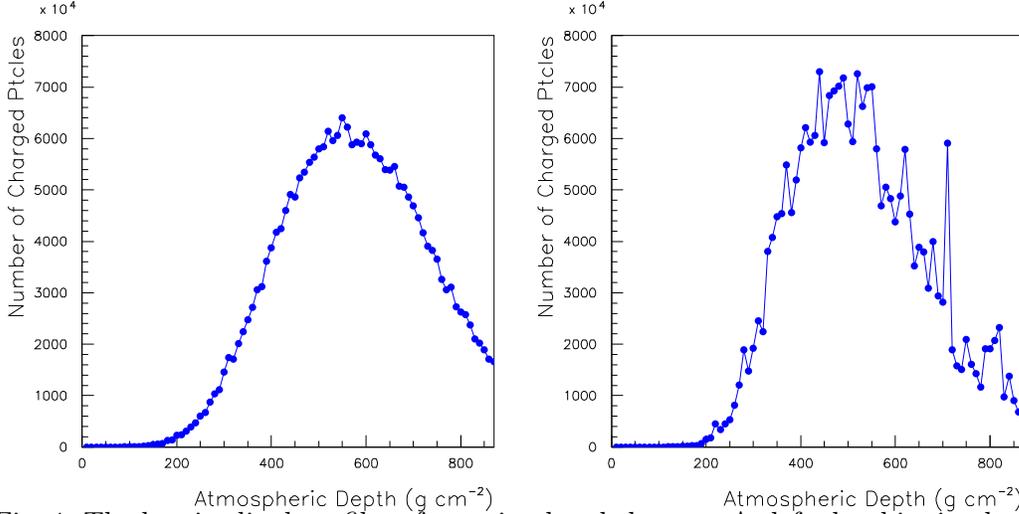}}
\end{center}
\vspace{-0.8cm}
\caption{The longitudinal profiles of two simulated showers.
At left the thinning level is $10^{-5.5}$ of primary energy,
and at right $10^{-3.5}$.}
\label{fig:exprof}
\end{figure}

When considering gross quantities, like the depth of shower maximum,
it is possible to run the simulation codes
with very severe thinning, and still obtain results of adequate quality.
This means that many showers can be generated, over a multidimensional
grid of primary parameters and shower models, within an acceptable
computing time.
The thinning process leads to longitudinal profiles which have large
non-statistical fluctuations.
The magnitude of these fluctuations increases with the severity of
the thinning; see Figure~\ref{fig:exprof}.

To recover the depth of shower maximum from ``noisy'' thinned
profiles it is customary to fit them to an empirical cascade
shape function.
This is also necessary when analyzing experimental data
as the quality is often poor.
The following function was introduced by Gaisser and
Hillas~\cite{GaisserHillas77} as a ``simple analytic parameterization''
of the longitudinal profile of air showers:

\begin{equation}
N(X) = \nm \left(\frac{X}{X_{max}}\right)
       ^{X_{max}/\lambda}
       \exp{(X_{max}-X)/\lambda},
\label{eqn:gh_func1}
\end{equation}

\noindent where $X$ is the atmospheric depth (in g~cm$^{-2}$), \nm\
is the number of cascade particles at shower maximum, \xm\ is the
depth of maximum, and $\lambda$ is a characteristic length parameter
(in the above reference fixed at 70~g~cm$^{-2}$).
This is a Gamma Function, a form which naturally arises in cascade
theory, and assumes that the first interaction is at $X=0$.

A fourth parameter $X_0$ is often introduced into Equation~\ref{eqn:gh_func1},
ostensibly to allow for a variable first interaction point,

\begin{equation}
N(X) = \nm \left(\frac{X-X_0}{X_{max}-X_0}\right)
       ^{(X_{max}-X_0)/\lambda}
       \exp{(X_{max}-X)/\lambda}.
\label{eqn:gh_func2}
\end{equation}

\noindent This seems somewhat inelegant as varying
$X_0$ does not correspond to a translation along the $X$ axis, unless
\xm\ is also changed.
The following (equivalent) form is physically clearer,
where $X_{rise}$ is the distance from first interaction to shower maximum,

\begin{equation}
N(X) = \nm \left(\frac{X-X_0}{X_{rise}}\right)
       ^{X_{rise}/\lambda}
       \exp{(X_{rise}+X_0-X)/\lambda}.
\label{eqn:gh_func3}
\end{equation}

In practice, when fitting simulated hadronic cascade profiles
to either Equation~\ref{eqn:gh_func2} or~\ref{eqn:gh_func3}, it turns out
that $X_0$ correlates poorly with the actual depth
of first interaction, and frequently takes unphysical negative values.
Experiments were made performing the fit with $X_0$ fixed at the actual
depth of first interaction $X_1$.
This produces a significantly poorer goodness of fit, and
reduces the \xm\ results by $\leq10$~g~cm$^{-2}$.
The choice of Equation~\ref{eqn:gh_func2} or~\ref{eqn:gh_func3} also
influences the \xm\ results.
For the remainder of this paper, in the interests of compatibility
with other published results, Equation~\ref{eqn:gh_func2} was used
with all four parameters free.
Note that $X_0$ is best regarded as simply an additional arbitrary
shape parameter.

To determine an appropriate thinning level for this study, sets of
500 proton showers were generated and fit,
at each of 5 thinning levels; $10^{-3.5}$, $10^{-4.0}$, $10^{-4.5}$,
$10^{-5.0}$ and $10^{-5.5}$.
Taking the $10^{-5.5}$ thinned distributions
as reference, a Kolmogorov test\footnote{
The test as implemented in the HBOOK function HDIFF~\cite{HBOOK_manual}
was used.}
was performed on each of the more heavily thinned sets, and for each of the
fit parameters.
This is a statistical test of the compatibility in shape between two
histograms --- it yields the probability that they are drawn
from the same parent distribution.
All the fit parameters returned a high probability at thinning
levels of $10^{-5.0}$ and $10^{-4.5}$, i.e.\ the results are
indistinguishable within the statistics of a 500 event set.
\xm\ itself is extraordinarily robust, remaining unbiased
even at $10^{-3.5}$ thinning.
To be conservative a value of $10^{-4.5}$ was selected for the main study.

The compatibility of the parameter distributions was also checked
when varying the zenith angle of the primary from 0 to 60~deg.
\xm\ was unaffected, but interestingly \nm\ showed a small
systematic {\it increase} with angle~\cite{GAP-98-035}.

%%%%%%%%%%%%%%%%%%%%%%%%%%%%%%%%%%%%%%%%%%%%%%%%%%%%%%%%%%%%%%%%%%%%%%%%%%%%%%%
\section{\xm\ Results}
\label{sec:emod}

%%%%%%%%%%%
\subsection{Mean value of \xm}

\begin{figure}
\begin{center}
\resizebox{\textwidth}{!}{\includegraphics{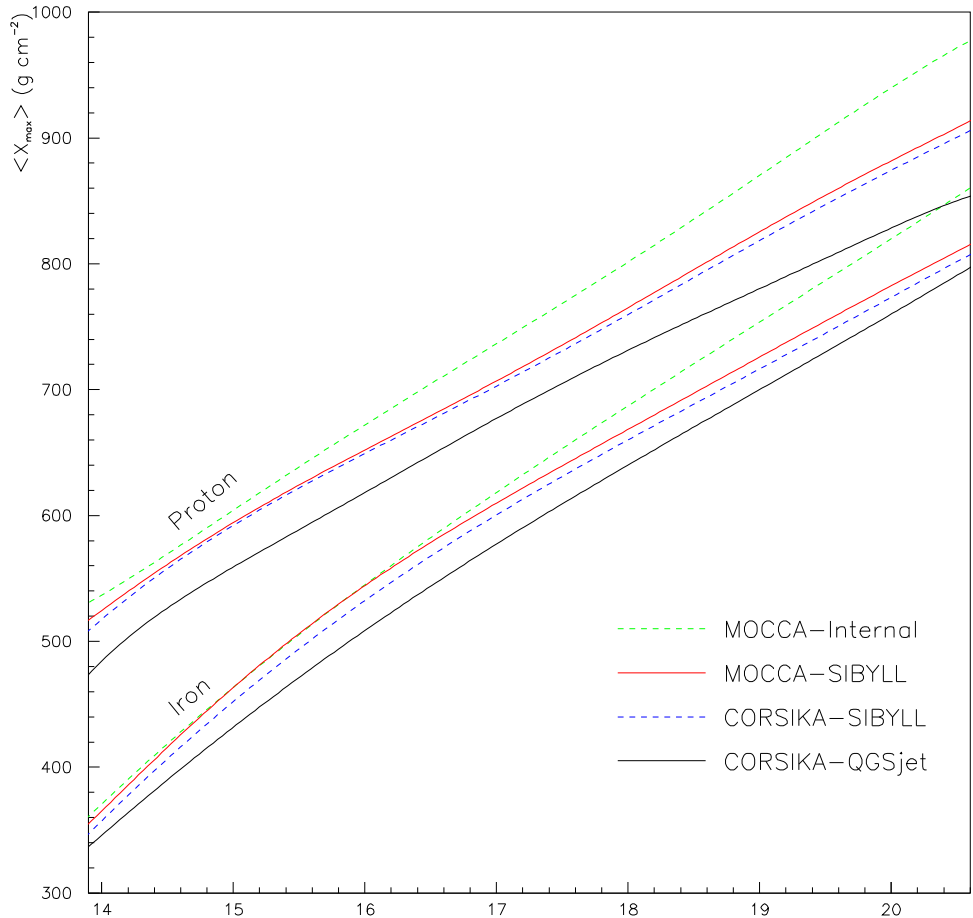}}
\end{center}
\vspace{-0.8cm}
\caption{The mean depth of shower maximum versus energy for
proton and iron primaries, and
four program-frame / interaction-model combinations.}
\label{fig:xmax}
\end{figure}

For the main study sets of 500 events were run at 14 half decade
energy steps between $10^{14}$ and $10^{20.5}$~eV, with the 4
combinations of framework program and high energy hadronic interaction
model given in Table~\ref{tab:simprogs}.
Sets were generated with both proton and iron nucleus primaries.
This gives $500 \times 14 \times 4 \times 2 = 56,000$ showers.
The showers were run at a thinning level of $10^{-4.5}$ of primary
energy, and a zenith angle of 45~deg.
The resulting profiles were fit to Equation~\ref{eqn:gh_func2}.
Figure~\ref{fig:xmax} shows the mean value
of \xm\ plotted against energy over the complete range;
numerical values are given in Table~3.
% \ref{tab:xmax}. Gives me the wrong number if use \ref correctly!
\mocsib\ and \corsib\ proton results are in good agreement,
and the iron results are also close.
This is very encouraging --- the framework programs are complex
and entirely independent --- nevertheless they produce the same result.
At higher energies the older \mocspl\ model diverges
strongly to deeper \xm.
\corqgs\ produces much shallower results than \sibyll\ at all energies;
so much so that at $10^{20.5}$~eV \mocspl\ iron is equal to \corqgs\ proton.

\begin{figure}
\begin{center}
\resizebox{\textwidth}{!}{\includegraphics{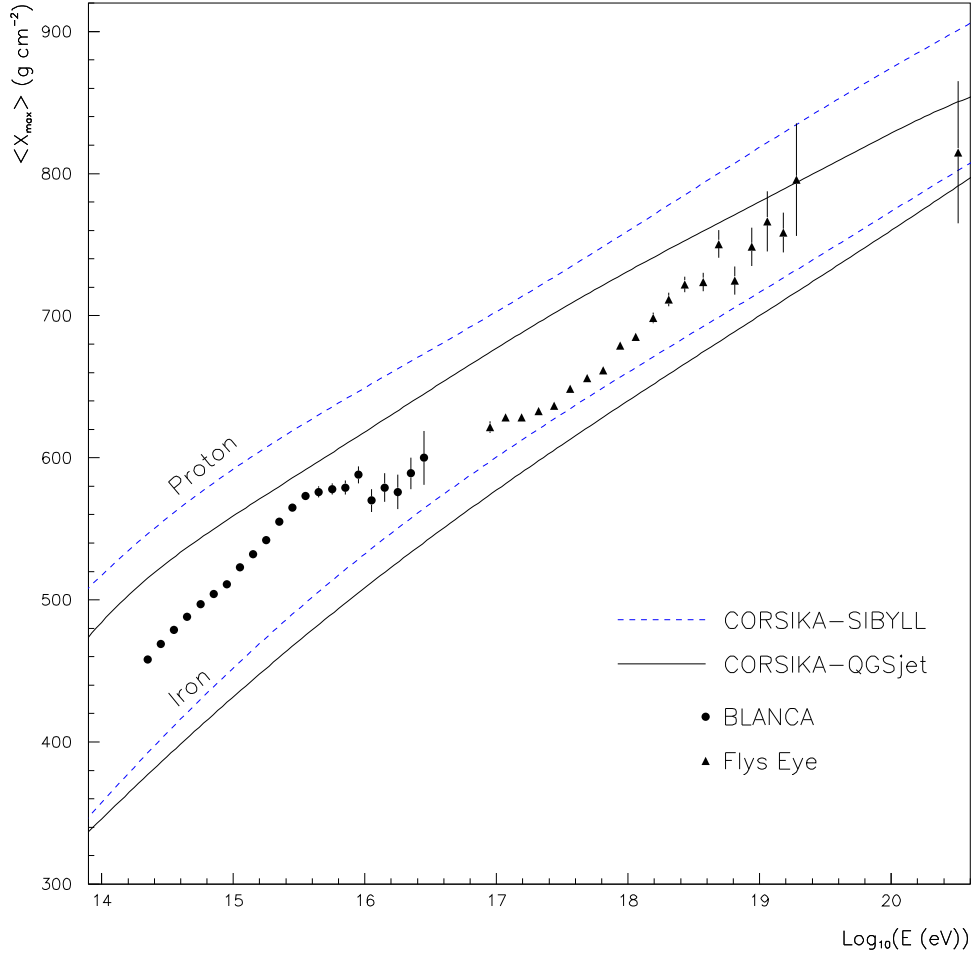}}
\end{center}
\caption{A Comparison of experimental mean \xm\ data with
simulations using two high energy hadronic interaction models.}
\label{fig:compex_mean}
\end{figure}

Figure~\ref{fig:compex_mean} shows a comparison between published
mean \xm\ data from two experiments and the \corsika\ calculations.
The data for $E<10^{17}$~eV are from the BLANCA experiment~\cite{blanca}, and
for $E>10^{17}$~eV from the Fly's Eye~\cite{Bird93_1}\footnote{
These points are used rather than the newer ones in~\cite{abuzayyad99} since
they have a much wider energy range, and their quoted errors are comparable,
or better.}.
The Fly's Eye data contains a small un-corrected experimental bias,
the removal of which would shift the lower energy points higher
in the atmosphere by around 20~g~cm$^{-2}$~\cite{gaisser93}.
Both \sibyll\ and \qgsjet\ are consistent with the data,
in that the value remains within the proton-iron bounds.
Also the general trends in composition versus energy are the same under the
two models, although the absolute value and size of the changes
differ.
There is some evidence at $\sim 10^{16}$~eV that \qgsjet\ is
a more realistic model than \sibyll~\cite{blanca,antoni99}; the extrapolation
to the highest energies in both models must be regarded as tentative at best.

%%%%%%%%%%%
\subsection{Influence of the First Interaction on \xm}
\label{sec:fint}

Why do the results shown in Figure~\ref{fig:xmax} differ between the models?
The proton-air cross sections used by \sibyll\ and \qgsjet\ are sufficiently
similar that the mean free paths differ by $<5$~g~cm$^{-2}$ over
the complete energy range.
This is to be compared to the 30--50~g~cm$^{-2}$ difference
in mean \xm.

\begin{figure}[t]
\begin{center}
\resizebox{0.5\textwidth}{!}{\includegraphics{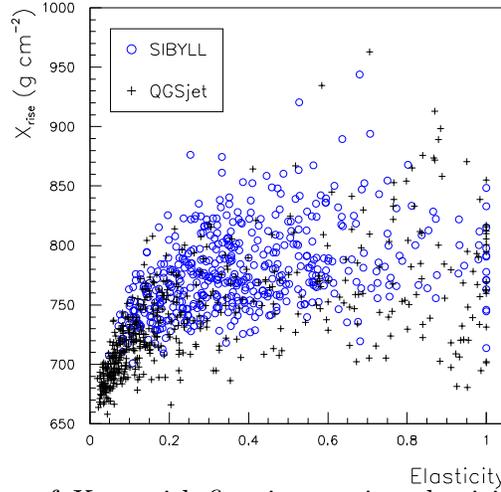}}
\end{center}
\vspace{-0.8cm}
\caption{The correlation of \xr\ with first interaction elasticity
for primary protons at $10^{19}$~eV.}
\label{fig:xrise_vs_elas}
\end{figure}

When investigating the way in which the first interaction controls
\xm\ it is natural to subtract
out the position of the first interaction; $X_{rise} = X_{max} - X_1$.
Elasticity is defined as the energy fraction of the most energetic secondary.
Normally a large fraction of the primary energy continues in the
form of a ``leading nucleon'' and the remainder is split between
many secondary pions and nucleons.
Figure~\ref{fig:xrise_vs_elas} shows \xr\ versus first interaction
elasticity at $10^{19}$~eV.
For events where the first interaction is catastrophic (small elasticity),
the resulting shower takes fewer generations to reach
maximum, and the correlation is strong.
As elasticity becomes larger, the first interaction is no longer the
controlling factor, and the relationship weakens.
Interestingly, both models exhibit approximately the same correlation
between elasticity and \xr.

\begin{figure}[t]
\begin{center}
\resizebox{\textwidth}{!}{\includegraphics{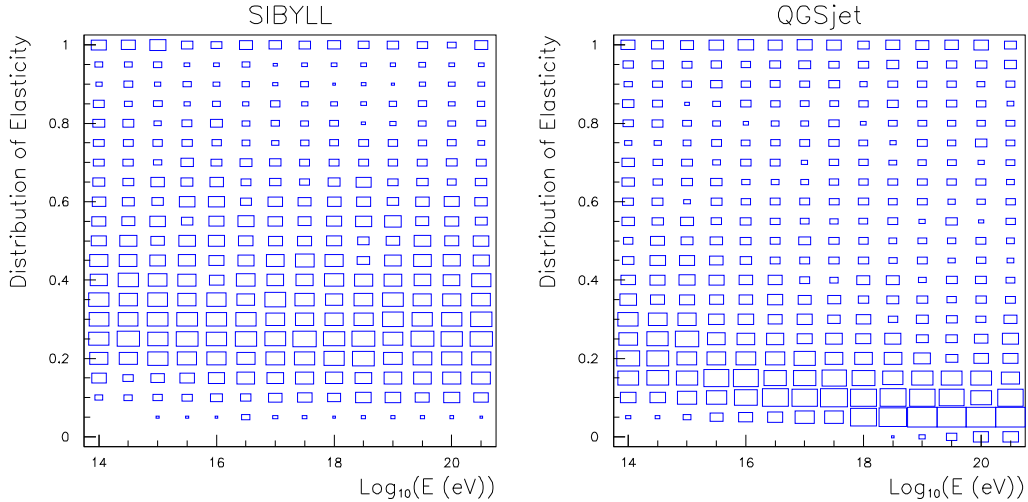}}
\end{center}
\vspace{-0.8cm}
\caption{Elasticity of proton-air interactions versus energy for two
models.
The \emph{side length} of the boxes is proportional to the bin contents.}
\label{fig:elas_dist}
\end{figure}

Figure~\ref{fig:elas_dist} shows the elasticity distributions
versus energy.
The reason the models produce different mean \xm\ values is
primarily because of their different elasticity distributions;
\qgsjet\ produces many more ``hard'' events, which lead to less deeply
penetrating showers.
\sibyll\ has rather constant behaviour versus energy, while \qgsjet\
is a more radical model, showing a stronger change; this is
why the corresponding mean \xm\ results diverge with increasing energy.

Hadronic interactions models are complex and esoteric,
with many parameters which can potentially be compared.
The significance of Figure~\ref{fig:xrise_vs_elas} is the
realization that, for calculations of \xm\ at least, the
most important characteristic is also a very simple one:
how much of the primary energy is expended in the first interaction?

%%%%%%%%%%%
\subsection{Distribution of \xm}

\begin{figure}[t!!!!]
\begin{center}
\resizebox{\textwidth}{!}{\includegraphics{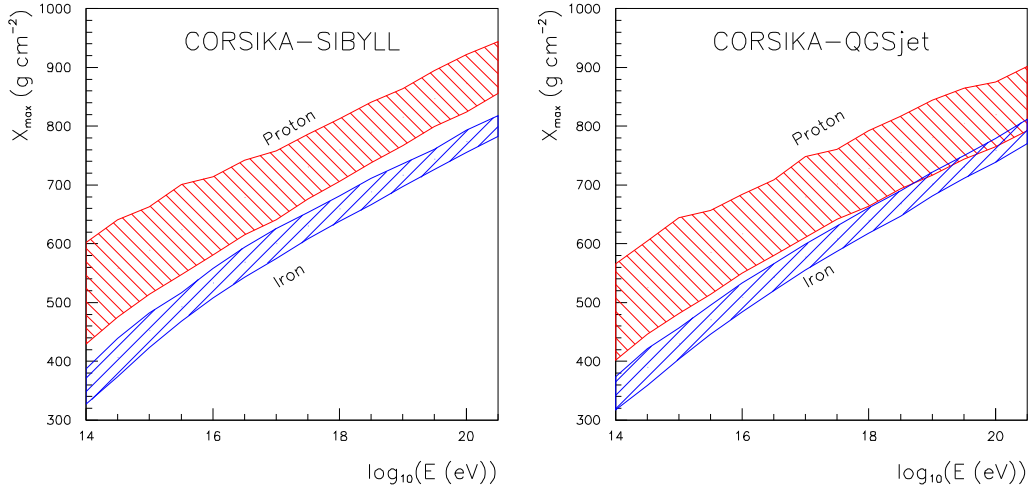}}
\end{center}
\caption{The distribution of \xm\ versus energy shown as a band
containing the central 68\% of the distribution.}
\label{fig:xmax_bands}
\end{figure}

In Figure~\ref{fig:xmax} it can be seen that \qgsjet\ predicts
that the difference between proton and iron mean \xm\ decreases
significantly with energy.
However, the fluctuations do not decrease correspondingly, and
the proton and iron distributions overlap to an increasing
extent.
If this model is correct, greater experimental statistics
would be required to determine the mean composition with given
accuracy.
The situation is illustrated in Figure~\ref{fig:xmax_bands}
where the bands contain 68\% of the data (i.e.\ spanning
the 16\% and 84\% points of the integral distribution).
\corsib\ shows stronger separation improving with increasing energy,
while \corqgs\ has weaker separation degrading with energy.

Proton primaries are deflected less by magnetic fields than
more highly charged particles of the same total energy.
It has been suggested that attempts to locate the origin
of the highest energy cosmic rays, by studying their arrival
directions, could be enhanced by making cuts on composition
sensitive parameters, to increase the fraction of protons in the data sample.
This would clearly work much less well if \qgsjet\ is a more realistic model
than \sibyll.

For proton primaries the distribution of \xm\ is strongly asymmetric,
with a tail to deep \xm.
This is presumably the result of fluctuations in the first interaction
point, and is therefore connected to the proton-air cross section, which
is a quantity of fundamental interest.
Earlier simulations have suggested that,

\begin{equation}
\Lambda = c \: \lambda_{p-air},
\end{equation}

where $\Lambda$ is the exponential slope, or ``decrement'', of the
trailing edge of the \xm\ distribution, $\lambda_{p-air}$ is the
proton-air mean free path, and $c$ is a constant of proportionality
with a value between 1.2 and 1.6 dependent on hadronic interaction
model~\cite{Ellsworth82}.

\begin{figure}[t]
\begin{center}
\resizebox{0.5\textwidth}{!}{\includegraphics{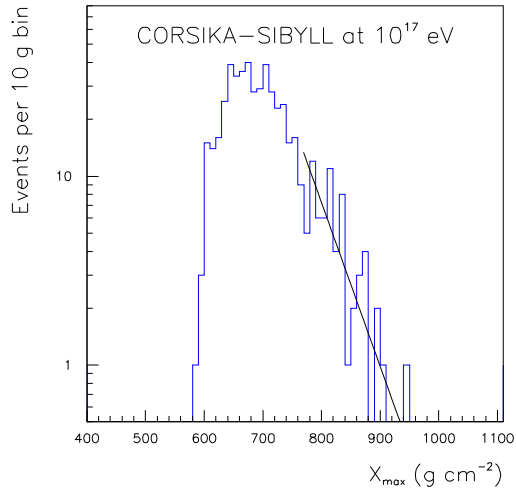}}
\end{center}
\caption{An example \xm\ distribution with exponential trailing edge fit.}
\label{fig:xmax_dist_fit}
\end{figure}

\begin{figure}[t]
\begin{center}
\resizebox{0.5\textwidth}{!}{\includegraphics{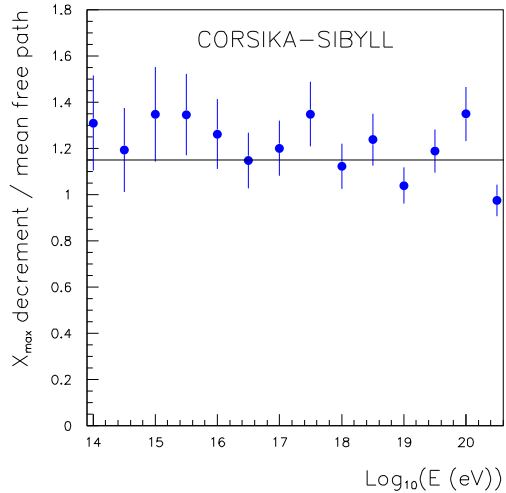}}
\end{center}
\caption{Ratio $c$ plotted versus energy.
%of \xm\ decrement to proton-air mean free path, plotted versus energy.}
%Assuming energy independence, 
The most probable value is shown as a horizontal
line.}
\label{fig:backedge}
\end{figure}

The \xm\ distributions were fit to an exponential starting 
at 100~g~cm$^{-2}$ beyond the peak.
An example distribution, with the fit, is shown in
Figure~\ref{fig:xmax_dist_fit}.
To avoid biasing the results it is essential to use a maximum
likelihood algorithm since the bins on the far tail necessarily
contain few events.
Figure~\ref{fig:backedge} shows the value of the ratio $c$,
plotted versus energy, for one of the models.
Testing each set of results against the hypothesis of energy
independence yields the values given in Table~\ref{tab:backedge}.
With the available statistics, the reduced $\chi^2$ numbers show
little evidence for energy dependence.
The \sibyll\ based models give values close to $1.15$, while
the other two are around $1.30$; this difference appears
to have significance.

\begin{table}
\begin{tabular}{|l|l|l|}
\hline
Model & Ratio $c$ & $\chi^2/ndf$ \\
\hline
\mocspl\ & $1.32 \pm 0.03$ & 0.77 \\
\mocsib\ & $1.16 \pm 0.03$ & 1.24 \\
\corsib\ & $1.15 \pm 0.03$ & 1.38 \\
\corqgs\ & $1.30 \pm 0.04$ & 0.43 \\
\hline
\end{tabular}
\label{tab:backedge}
\caption{Decrement / mean-free-path data.}
\end{table}

%%%%%%%%%%%%%%%%%%%%%%%%%%%%%%%%%%%%%%%%%%%%%%%%%%%%%%%%%%%%%%%%%%%%%%%%%%%%%%%
\section{Conclusions}

When running shower simulations to study \xm\ it is better
to generate heavily thinned showers, with
explicit low energy hadronic and electromagnetic cascades, than to rely
on analytic approximations for these parts of the calculation.
The latter has frequently been done in the past, leading to concerns
that the results are biased to an unknown extent.
When working with an explicit, but thinned, cascade simulation it is
possible to determine an appropriate thinning level empirically,
by comparing against more lightly thinned results.

Carefully calculated \xm\ results have been presented,
over a wide energy range, for proton \& iron primaries, using four
combinations of framework program and high energy hadronic interaction
model.
It is hoped that these will be of use for future comparisons with
experimental data, and with other simulation results.

The way in which the first interaction controls \xm\ has been
investigated.
The influence is strong --- if one were to use model $A$ for the first few
interactions, and model $B$ thereafter, the mean \xm\ results would be
close to using model $A$ throughout.
\qgsjet\ predicts that the separation between proton and iron \xm\
declines at the highest energies.
If this is true it is unfortunate from an experimental perspective.

It would be very useful if a common reference set of showers were
made available by the authors of new, or modified, hadronic interaction models.
For the purposes of longitudinal profile comparison the set used here
seems adequate; the raw and processed output is available
online~\cite{data}.

The Fermilab computing department are thanked for the use of
their machines.

%%%%%%%%%%%%%%%%%%%%%%%%%%%%%%%%%%%%%%%%%%%%%%%%%%%%%%%%%%%%%%%%%%%%%%%%%%%%%%%
\begin{table}
\begin{tabular}{|l|l|c|c|c|c|}
\hline
Primary & $\log_{10}(E (eV))$ & \multicolumn{4}{c|}{
$<\!X_{max}\!>$, $\sigma X_{max}$ (g~cm$^{-2}$)} \\
\cline{3-6}
 & & \multicolumn{2}{c|}{MOCCA} & \multicolumn{2}{c|}{CORSIKA} \\
\cline{3-6}
&  & Internal & SIBYLL & SIBYLL & QGSjet \\
\hline Proton 
& 14.0 & 537 , 96 & 525 , 91 & 517 , 97 & 484 , 96 \\
& 14.5 & 570 , 86 & 559 , 92 & 560 , 99 & 525 , 92 \\
& 15.0 & 605 , 78 & 597 , 86 & 589 , 97 & 560 , 88 \\
& 15.5 & 637 , 72 & 625 , 81 & 621 , 84 & 587 , 76 \\
& 16.0 & 674 , 71 & 651 , 72 & 651 , 81 & 618 , 72 \\
& 16.5 & 704 , 62 & 676 , 59 & 679 , 70 & 646 , 70 \\
& 17.0 & 738 , 60 & 713 , 65 & 699 , 63 & 679 , 75 \\
& 17.5 & 767 , 54 & 733 , 57 & 729 , 63 & 705 , 68 \\
& 18.0 & 802 , 53 & 765 , 60 & 762 , 59 & 728 , 69 \\
& 18.5 & 839 , 50 & 793 , 53 & 790 , 62 & 755 , 67 \\
& 19.0 & 867 , 44 & 829 , 55 & 817 , 52 & 779 , 68 \\
& 19.5 & 908 , 60 & 852 , 49 & 847 , 52 & 804 , 68 \\
& 20.0 & 940 , 50 & 883 , 53 & 875 , 54 & 825 , 59 \\
& 20.5 & 972 , 42 & 908 , 47 & 900 , 45 & 849 , 59 \\
\hline Iron 					    
& 14.0 & 370 , 36 & 366 , 38 & 357 , 32 & 346 , 32 \\
& 14.5 & 420 , 34 & 414 , 35 & 407 , 33 & 390 , 30 \\
& 15.0 & 461 , 34 & 464 , 34 & 452 , 31 & 432 , 29 \\
& 15.5 & 507 , 32 & 508 , 34 & 493 , 28 & 471 , 26 \\
& 16.0 & 545 , 29 & 542 , 31 & 533 , 27 & 509 , 25 \\
& 16.5 & 582 , 25 & 577 , 27 & 568 , 27 & 544 , 26 \\
& 17.0 & 619 , 23 & 610 , 29 & 600 , 25 & 577 , 24 \\
& 17.5 & 652 , 22 & 641 , 26 & 631 , 25 & 609 , 22 \\
& 18.0 & 688 , 21 & 668 , 24 & 660 , 25 & 640 , 24 \\
& 18.5 & 720 , 18 & 697 , 21 & 689 , 24 & 669 , 23 \\
& 19.0 & 754 , 18 & 725 , 22 & 717 , 21 & 701 , 20 \\
& 19.5 & 787 , 17 & 754 , 21 & 744 , 20 & 730 , 21 \\
& 20.0 & 820 , 16 & 783 , 19 & 774 , 20 & 759 , 21 \\
& 20.5 & 853 , 14 & 809 , 19 & 802 , 20 & 791 , 22 \\
\hline
\end{tabular}
\label{tab:xmax}
\caption{Mean and standard deviation values of \xm\
for proton and iron primaries, and four program-frame / interaction-model
combinations.
Each pair of numbers comes from a 500 shower set.}
\end{table}

%%%%%%%%%%%%%%%%%%%%%%%%%%%%%%%%%%%%%%%%%%%%%%%%%%%%%%%%%%%%%%%%%%%%%%%%%%%%%%%

\end{document}